\documentclass[cup6a]{cupbook}
\usepackage{rob-999}

\newcommand{\prd}{\textit{Phys Rev D}}
\newcommand{\apj}{\textit{Astrophys J}}

\newcommand{\mnras}{\textit{MNRAS}}
\newcommand{\apjl}{\textit{Astrophys J Lett}}
\newcommand{\apjs}{\textit{Astrophys J Supplement}}

\def\bea{\begin{eqnarray}}
\def\ena{\end{eqnarray}}

\usepackage{amssymb}
\usepackage{amsmath}
\usepackage{graphicx}

\def\n{{\bf n}}

\def\P{{\bf P}}

\def\bi{\begin{itemize}}
\def\ei{\end{itemize}}

\def\n2{N$_{2}$}
\def\4he{$^{4}$He}
\def\cm3{cm$^3$}

\def\bea{\begin{eqnarray}}
\def\ena{\end{eqnarray}}

\newcommand{\arcdeg}{^\circ}

\def\beq{\begin{equation}}
\def\eeq{\end{equation}}

\title[CMB Polarization Tests of the Inflationary Paradigm]
     {An ``Ultrasonic Image" of the Embryonic Universe} \author{Brian G. Keating} \date{17 May 2008}

\begin{document}
\pagenumbering{roman}
\maketitle
\pagenumbering{arabic}

\chapter[An ``Ultrasonic" Image of the Embryonic Universe]
{An ``Ultrasonic" Image of the Embryonic Universe: CMB
Polarization Tests of the Inflationary Paradigm}

\section{Introduction}
Why was there a Big Bang? Why is the universe not featureless and
barren? Why are there fluctuations in the cosmic microwave
background? In stark contrast to the convincingly answered ``what"
questions of cosmology (e.g. What is the age of the universe?,
What is the geometry of the universe?), these ``why" questions
may instead evoke a sense of disillusionment. Is it possible that
cosmology's ``triumphs"---its answers to the ``what" questions---are
frustratingly inadequate, or worse, incomplete?

However, what if the ``why" questions provide tantalizing hints of
the ultimate origins of the universe? Then instead of crisis, we
encounter an amazing opportunity---one that might provide answers
to the most enigmatic question of all: How did the universe begin?

Inflation \cite{guth81} is a daring paradigm with the promise to
solve many of these mysteries. It has entered its third decade of
successfully confronting observational evidence and emerged as
cosmology's theoretical touchstone. Despite its many successes,
inflation remains unproven. While skeptics must resort to
increasingly finely tuned attacks \cite{khoury01,magueijo03},
inflation's proponents can only cite circumstantial evidence in
its favor \cite{turner02}. However, a conclusive detection of a
primordial gravitational wave background (GWB) from inflation
would be ``the smoking gun" \cite{kamkos1998}. No other known
cosmological mechanism mimics the GWB's imprint on the cosmic
microwave background (CMB).

New technological innovations poise cosmology at the threshold of
an exhilarating era---one in which future CMB data will winnow
down the seemingly boundless ``zoo" of cosmological models and
test the hypothesis that an inflationary expansion of the universe
took place in its first moments.

Inflation's unique imprint on CMB polarization has generated
considerable attention from US science policy advisors
\cite{nrc2003,nasmckee,nasurry,hepap}, who have all
enthusiastically recommended measuring CMB polarization. The
reason for this excitement is clear: inflation explains a host of
critical cosmological observations, and CMB polarization is the
most promising, and perhaps only, way to glimpse the GWB.

This chapter describes how the Cosmic Gravitational Wave
Background induces a specific type of CMB polarization and
describes the first experiment dedicated to testing this
most-promising signature of inflation. This experiment, the
Background Imaging of Cosmic Extragalactic Polarization (BICEP)
project, has recently embarked on its third observing season. We
show preliminary data from the BICEP's first season obtained with
a novel polarization modulation mechanism called the ``Faraday
Rotation Modulator". Our discussion ends with a description of
exciting new technology with the potential to probe inflation down
to the ultimate cosmological limit.

\section{The inflationary universe}

The CMB has historically been \emph{the} tool to appraise
inflationary cosmology. This is not surprising since the CMB is
the earliest electromagnetic ``snapshot" of the universe, a mere
380,000 years after the Big Bang. As such it probes the universe
in a particularly pristine state---before gravitational and
electromagnetic processing. Because gravity is the weakest of the
four fundamental-forces, gravitational radiation (i.e. the GWB)
probes much farther back: to $\simeq 10^{-38}$ seconds after the
Big Bang ($10^6 t_{Pl}$ in Planck units).  The GWB encodes the
cosmological conditions prevailing at $10^{16}\, \rm{GeV}$ energy
scales. In contrast, the CMB encodes the physical conditions of
the universe when radiation decoupled from matter at energy scales
corresponding to 0.3 eV at $t \simeq 10^{56}t_{Pl}$. As
experimentalists, we can exploit the primacy of the CMB by using
the CMB's surface of last scattering as a ``film" to ``expose" the
GWB---primordial reverberations in spacetime itself. Doing so
will provide a ``baby picture" of the infant universe; an
``ultrasonic" image of the embryonic universe!

\subsection{Quantum fluctuations in the inflationary universe}
Inflation posits the existence of a new scalar field (the
\emph{inflaton}) and specifies an action-potential leading to
equations of motion. Quantizing the inflaton field causes the
production of perturbations (zero point fluctuations)
\cite{borner03}. While the inflaton's particle counterpart is
unknown, its dynamics as a quantum field have dramatic
observational ramifications \cite{kinney97}. All viable
cosmological theories predict a spectrum of scalar (or energy
density) perturbations that can then be tested against CMB
temperature anisotropy measurements. Inflation predicts the
spectrum of scalar perturbations, and additionally predicts tensor
perturbations (i.e. the GWB). Inflation's unique prediction is
the GWB, which is parameterized by the tensor-to-scalar ratio,
$r$. An unambiguous detection of $r$ will reveal both the epoch of
inflation and its energy scale \cite{kamkos1998}. If, as theorists
have speculated \cite{kamkosowsky99}, inflation is related to
Grand Unified Theories (GUT), then a detection of the GWB also
will probe physics at energy scales one trillion times higher than
particle accelerators such as the Large Hadron Collider
\cite{lhc}.

Both energy density fluctuations (scalar perturbations) and
gravitational radiation (tensor perturbations; the GWB) produce
CMB polarization. The two types of perturbations are related in
all inflation models, since both are generated by quantum
fluctuations of the same scalar field, the inflaton
\cite{liddlelyth2000}. The relationship between CMB polarization
produced by scalars and tensors will provide a powerful
consistency check on inflation when the GWB is detected. Similar
relations, using recent detections of the \emph{scalar}
perturbation spectrum's departure from ``scale invariance"
\cite{wmap3} (primarily using CMB temperature anisotropy), have
led to claims of ``detection" of inflation, at least in the
popular press \cite{guthpostwmap}. Furthermore, NASA's Wilkinson
Microwave Anisotropy Probe (WMAP) showed an anti-correlation
between temperature and polarization at large angular scales,
providing additional, albeit circumstantial, evidence in favor of
inflation \cite{spergelzaldarriaga1997, page2006}.

The ultimate test of inflation requires a measurement of the
tensor power spectrum itself, not only the predicted
temperature-polarization correlation or the properties of the
\emph{scalar} power spectrum. Given a very modest set of external,
non-inflation-specific parameters, including a simple cosmological
chronology (specifying that inflation was followed by radiation
domination, subsequently followed by matter domination),
inflationary models can precisely predict the spatial correlations
imprinted on the polarization of the CMB by the GWB,
making it truly ``the smoking gun."

\subsection{The gravitational wave background: shaking up the
CMB}

Scalar metric-perturbations have no handedness, and are therefore
said to be ``parity invariant." While the GWB produces both
temperature and polarization perturbations, the temperature
perturbations are primarily associated with the change in
potential energy induced by the gravitational waves, whereas the
tensor-induced polarization perturbations are associated with
spacetime stress and strain. The parity-violating polarization
signature exists only if cosmological gravitational waves exist,
as first demonstrated by Polnarev \cite{polnarev85}. As we will
show, the amplitude of the polarization is determined by the
energy scale of inflation, and its angular/spatial correlation
structure is determined nearly exclusively by the expansion of the
universe. While the energy scale (and, quite frankly, even
\emph{the existence}) of the inflaton is unknown, the
post-inflation expansion history of the universe is
\emph{extremely} well understood. This is quite fortuitous for
experimentalists hunting for the GWB as it dramatically restricts
the range of our prey!

The separation between the inflationary and standard hot Big Bang
dependencies is yet another manifestation of the interplay between
inflation's quantum mechanical aspects and the Big Bang
cosmology's classical dynamics. Although the inflationary
perturbations are quantum mechanical in origin, they are small
enough to be treated using linearized classical general relativity
(the so-called WKB semi-classical approximation). So while the
tensor-to-scalar ratio will probe quantum cosmology, the
(classical) evolution of the scale factor allows for a precise
prediction of the GWB's angular correlation imprint on CMB
polarization. This separability, into a classical part (sensitive
to the background evolution of spacetime) and a small,
perturbative quantum component, makes the CMB's curl-mode
polarization the most robust probe of inflation.

\subsection{Observations and challenges}
The inflationary model has revolutionized cosmology. Inflation
solves the ``horizon problem"---reconciling observations that
show that regions of the universe have identical CMB temperatures
(to a part in $10^5$) by providing a causal mechanism for these
regions to attain thermal equilibrium 380,000 years after the Big
Bang. Inflation solves the horizon problem via an exponential,
accelerating expansion of the universe at early times, prior to
the ``ordinary" Hubble-Friedmann expansion observed today. This
rendered the entire observable universe in causal contact
initially, and also accounts for the seemingly finely tuned
spatial flatness of the universe observed by CMB temperature
anisotropy experiments
\cite{deBernardis00,balbi00,pryke02,sievers03,spergel2003,goldstein03}.

Inflation also predicts a nearly scale-invariant spectrum of
scalar perturbations. That \emph{any} initial perturbations remain
after the universe expanded by a factor of $\sim e^{60}$ is
astonishing! Yet, surprisingly, the fluctuation level at the
surface of last scattering arises naturally \cite{liddlelyth2000}
in inflation as a consequence of parametric amplification, see
Section \ref{s:parametric_amplification}. The residual
fluctuations are observable in the CMB and indicate the epoch of
inflation and the amount of expansion (the duration of inflation).
This is inflation's solution to cosmology's ``smoothness problem,"
accounting for the small, but non-vanishing, level of
perturbations. Regrettably, neither flatness nor smoothness are
unique to inflation. Both have long histories, predating
inflation. Flatness was anticipated on quasi-anthropic
principles \cite{dicke65}, and the universe's near-smoothness was
predicted as the primordial matter power
spectrum \cite{harrison70,peebles70,zeldovich72}. Recent CMB and
galaxy cluster measurements \cite{eisenstein05,cole05} have
detected perturbations (possibly) resulting from
phase-synchronized ``quantum noise" (zero-point oscillations in
the inflaton). These scalar, mass/energy perturbations, combined
with the universe's spatial flatness, increase inflation's
credibility since the $e^{60}$-fold expansion producing flatness
\emph{should} have also destroyed all initial perturbations.
However, while there is abundant circumstantial evidence, there is
one unique prediction of inflation: the primordial GWB,
which produces an unmistakable imprint on the
polarization of the CMB.

\section{CMB polarization}

The CMB is specified by three characteristics: its spectrum, the
spatial distribution of its intensity (or temperature anisotropy),
and the spatial distribution of its polarization. All three
properties depend on fundamental cosmological parameters.
Additionally, since CMB photons travel through evolving structures
in the early universe on their way to our telescopes today, the
CMB is also a probe of cosmic structures along the line of sight,
which are, in some sense, ``foregrounds"---either emitting,
attenuating, or distorting the spatial and frequency power spectra
of the background.

Originally proposed by Rees \cite{rees68} as a consequence of an
anisotropically expanding universe, the polarization of the CMB
was unobserved for many decades. Although Rees' original model was
found to be untenable, it was later corrected by Basko and
Polnarev \cite{basko} in 1980. Nevertheless Rees' exploration
attracted the attention of
experimentalists \cite{caderni78,lubin79,nanos79} who initiated
observations to measure CMB polarization. The polarization of the
CMB, and its correlation with temperature anisotropy, was first
detected by DASI \cite{kovac02}. The race to discover the wispy
imprint of the GWB was on!

\subsection{Temperature anisotropy produced by the GWB}

When electrons in the primordial plasma prior to decoupling were
irradiated with CMB photons, polarization of the microwave
background was inevitable. Thomson scattering (low energy Compton
scattering) produces polarization whenever photons from an
anisotropic radiation field scatter off unbound electrons.
Anisotropy in the CMB radiation field was produced by either
mass/energy perturbations (over- and under-dense regions) or
gravitational waves. When the photon field is decomposed into
spherical harmonics, these two types of perturbation produce
anisotropy of the quadrupolar variety ($Y_{\ell,m}$ with $\ell =
2$). There are five harmonics with $\ell =2$, but only one of
these, with $m=0$, is azimuthally symmetric. The $Y_{2,m}$ with
$m=\pm2$ indicate that gravitational waves are spin-2 objects
\cite{huwhite97}. Gravitational waves ``shear" spacetime and
produce local violations of reflection, or parity, symmetry in the
CMB polarization field \cite{polnarev85,basko}.

The GWB produces CMB temperature anisotropy as well. However, the
temperature anisotropy is a scalar field on the celestial sphere
and is dominated by the acoustic oscillations of radiation and
matter, overwhelming the minute temperature anisotropy produced by
gravitational waves. The temperature anisotropy induced by the GWB
is also degenerate with other cosmological
parameters \cite{kamkossteb,selzal97} and essentially undetectable
at levels below the current WMAP3 limits, due to cosmic variance
\cite{knoxturner94}.

\subsection{Polarization anisotropy produced by the GWB}
\begin{figure}
\centering
\includegraphics[height=8cm,angle=90]{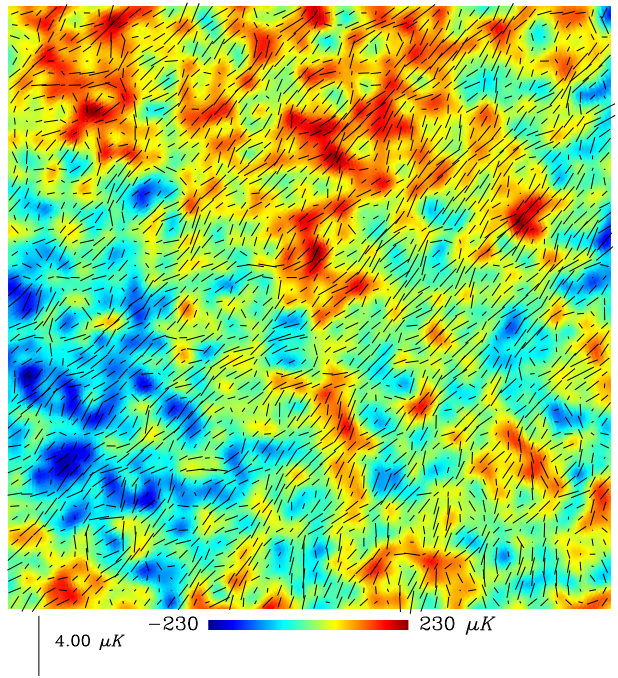}
\caption{A simulated noiseless map of CMB polarization and
temperature anisotropy. The simulation represents an $18\arcdeg
\times 18\arcdeg$ map of the CMB's temperature, E-mode, and B-mode
polarization. The temperature (gray scale) and E-mode polarization
reveal the classical cosmological parameters such as the mass
density, geometric curvature, and composition of the universe
(i.e. ``dark" versus ordinary matter). The B-mode, or ``curl,"
polarization is \emph{only} generated by primordial gravitational
waves and is indicated by regions where reflection symmetry is
locally violated. The vertical scale bar at the lower left
indicates polarization at the
$4~\mu$K level. The polarization vectors are the sum of E- and
B-mode polarization, but are dominated by E-mode polarization
(B-mode polarization is less than 0.1$\,\mu$K in this simulation,
corresponding to a tensor-to-scalar ratio $r\sim 0.1$). Figure
credit: Nathan Miller.} \label{f:polmaps}
\end{figure}

Fortunately, however, the CMB polarization's \emph{tensorial
nature} breaks the parameter degeneracy. Using an analog of
Helmholtz's vector calculus theorem valid for spin-2 fields on the
celestial sphere, CMB polarization maps, like Figure
\ref{f:polmaps}, can be decomposed into two scalar fields or
``modes" \cite{kamkossteb,selzal97}. The advantage of manipulating
two scalar fields, as opposed to one tensor field, is
self-evident.

\begin{figure}
\centering
\includegraphics[height=8cm]{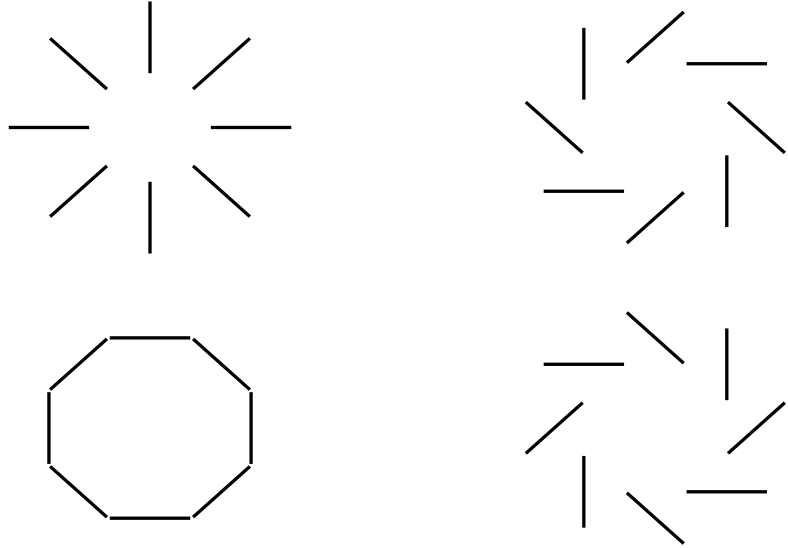}
\caption{Parity symmetric E-mode (or ``gradient-mode")
polarization patterns (left side), and B-mode (or ``curl-mode")
patterns (right side) in real space. Figure credit: Nathan
Miller.} \label{f:polmodes}
\end{figure}

One of the scalar fields is, essentially, the gradient of a
scalar-potential and is known as ``E-mode," or ``gradient-mode,"
polarization by analogy to the electric field. The E-mode
polarization is invariant under parity transformations. The second
component, called ``B-mode," or ``curl-mode," polarization is
analogous to the the curl of a vector-potential. If inflation
produced a sufficient amount of gravitational radiation, then
future maps of CMB polarization will be admixtures of both modes
(though the E-mode polarization will dominate by \emph{at least} a
factor of ten). For reference, simple one-dimensional maps of pure
E- and B-modes are shown in Figure \ref{f:polmodes}. Appraising the
behavior of the circular maps with respect to reflections across
the map's diameter reveals the symmetry of the underlying
polarization mode.

The pioneering work by Polnarev \cite{polnarev85} was the first to
identify a unique observational signature of gravitational waves;
one that would \emph{only} be manifest in the polarization of the
CMB. Polnarev's key insight was to recognize that asymmetric shear
induced by gravitational waves would induce a polarization pattern
significantly different from that produced by scalar
perturbations, such as those shown in the right-hand side of
Figure \ref{f:polmodes}. Polnarev predicted that gravitational
waves would be the only plausible source of parity violation on a
cosmological scale. In the more modern language of E- and B-modes,
this is equivalent to predicting the existence of B-modes
imprinted on the CMB by a primordial GWB.

\section{The origin of the gravitational wave background}

How can \emph{any} perturbations originating from quantum
fluctuations in the primordial inflaton field survive the
explosive expansion by a factor of $e^{60}$? After all, a hallmark
of inflation is that this expansion dilutes the curvature of the
universe from any primordial value to precisely flat. In fact,
since the GWB is a radiation background exactly like the CMB, the
subsequent expansion following inflation dilutes the GWB's energy
density by a factor of $a^4$, where $a$ is the cosmic scale
factor. Since the GWB energy density today is at least one billion
times smaller than the CMB's energy density, this means that it
was utterly insignificant for all times, including at decoupling.

This brings us to another potentially troubling ``why" question:
Why are gravitational waves expected to persist to
last-scattering, and leave a detectable imprint on the CMB, if the
very cosmological model predicting them produces an expansion
that should render them negligible? A hint at the answer to this
question comes from a rather unlikely source: a playground
swingset!

\subsection{Parametric resonance and amplification}
\label{s:parametric_amplification}

Parametric amplification is exemplified by an undamped pendulum of
length $L$ whose suspension point, $y$, is driven vertically $y=A
\cos{2\pi f t}$. To highlight the connection between (1) the
GWB, (2) the vertically driven pendulum,
and (3) the laser, we term the periodic driving force the ``pump."
Pump energy at frequency $f$ drives the pendulum into resonance
and can even amplify small random thermal vibrations of an
(initially stationary) pendulum into oscillation.

For simplicity, the angle between the pendulum and the vertical,
$\phi$, is taken to be small, and the equation of motion for $\phi$
becomes:

\beq \ddot{\phi}+(\omega_0^2 + \Omega^2 \cos 2\pi f t) \phi=0
\label{e:mathieu}\eeq where $\Omega^2 = (2 \pi f)^2 A/L$.

Equation \ref{e:mathieu} is known as \emph{Mathieu's equation}. In
Mathieu's equation, $\omega_0$ is the natural frequency of the
pendulum, and the \emph{parameter} $\Omega$ determines the resonant
behavior of the pendulum, leading to \emph{parametric resonance}.
While the equations of motion are linear with respect to $\phi$,
the effect of the pump is not additive (as it would be if the
suspension point were to be horizontally modulated), but rather
\emph{multiplicative}.

To find periodic solutions to Equation \ref{e:mathieu}, we construct the
following \emph{ansatz}:

\beq \phi = \varphi_+ e^{+i\pi ft} + \varphi_- e^{-i\pi
ft}\label{e:ansatz} \eeq

In general, both stable and unstable solutions of Equation
\ref{e:mathieu} can be obtained. So called ``resonance bands" are
separated by regions of stability where the amplitude of $\phi$ is
constant. Unstable solutions exponentially diverge (as $\phi
\simeq e^t$). Both types of solutions will be important in the
context of gravitational waves. Solving Equation \ref{e:mathieu} using
Equation \ref{e:ansatz} leads to \beq [\omega_o^2 - (\pi^2
f^2)]\varphi_\pm + \frac{\Omega^2}{2}\varphi_\mp=0
\label{e:solution}\eeq where third-harmonic generation effects
have been ignored. Stable solutions to the (two) Equations
\ref{e:solution} are non-trivial and solvable for $\varphi_\pm$
when the following self-consistency relation holds between the
pump amplitude, frequency, and natural frequency of the pendulum:
\beq \Omega^2 = 2[\omega_o^2 - \pi^2 f^2]
\label{critical_amp}\eeq

The first, or \emph{fundamental}, resonance condition for the pump
frequency is $f=\omega_0/\pi$, which implies that pumping at
\emph{twice} the pendulum's natural frequency defines the boundary
of incipient instability even if the pump amplitude is small
($|\phi(t)| \neq 0$ even as $\Omega, A \rightarrow 0$).

The pumping strategy mentioned above shares similar features with
a similar application on the playground. Assisted swinging on a
swingset is a resonant system with pump-power supplied by two
assistants, one at each of the two displacement maxima;
that is, with pump frequency $f=2\,\frac{\omega_0}{2\pi}$. When
this condition holds, $\Omega=0$ and only small amounts of pump
energy are required to maintain the swing's oscillatory behavior,
even in the presence of significant frictional damping (which has
been ignored here).

In fact, while less social (and more dangerous) than assisted
pumping (using two friends), vertical pumping employing parametric
resonance allows the rider to initiate resonance by themselves (by
raising and lowering their center of mass---alternately standing
and squatting on the swing). Surprisingly, the amplification of
small initial perturbations via parametric resonance provides a
fruitful analogy for the theory of gravitational wave
amplification.

\subsection{Parametric amplification of the GWB}

Gravitational waves have unique and fascinating cosmological
properties. While the contribution of these waves to the energy
density of the universe today is minuscule, parametric
amplification of these primordial quantum fluctuations of the
inflaton field causes the waves to grow large enough to become
potentially observable. If the imprint of these primordial
perturbations is observable, an understanding of the parametric
amplification process allows us to optimize our observational
requirements---regardless of the magnitude of the inflationary
GWB.

In our simplified cosmology, spacetime is smooth and flat. On top
of this background a tensor field, representing the GWB, is suffused. The metric of this spacetime
is obtained by solving the Einstein equations
$$G_{\alpha\beta}=8\pi G T_{\alpha\beta}$$
where $G$ is Newton's constant of universal gravitation. For empty
space the stress-energy tensor $T_{\alpha\beta}=0$, leading to

\beq ds^2 = a(\eta)^2[d\eta^2 -
(\delta_{\alpha\beta}+h_{\alpha\beta}) dx^\alpha dx^\beta]
\label{e:metric}\eeq where $\delta_{\alpha\beta}$ is the Kronecker
delta function, and $\eta$, the \emph{conformal time}, is related
to the (time-dependent) cosmological scale factor $a$ via
$d\eta=dt/a$. The (linearized) perturbation tensor is both
transverse-symmetric ($h_{\alpha\beta}=h_{\beta\alpha}$) and
traceless ($\sum_\alpha h_{\alpha\alpha}=0$). We seek solutions,
which are separable into a tensorial part and a scalar part, of
the following form

\beq h_{\alpha\beta} \equiv \sqrt{8\pi
G}\,\frac{\nu}{a}\,\epsilon_{\alpha\beta}\,e^{ik\eta}\label{e:h_eq_nu_div_a}\eeq
where $\epsilon_{\alpha\beta}$ is the gravitational wave
polarization tensor. The rank-two tensors $h_{\alpha\beta}$ and
$\epsilon_{\alpha\beta}$ are transverse and traceless, leading to
two independent polarization modes denoted ``$+"$ and ``$\times$."
Solving Einstein's equations yields wave equations for $\nu$, the
(scalar) amplitude of the set of equations $h_{\alpha\beta}$

\bea
\ddot{\nu} + (k^2 - \ddot{a}/a)\nu &=& 0\nonumber\\
\ddot{\nu} + (k^2 - U_{eff})\nu &=& 0 \label{e:gwbpara}
\label{e:gwb_mathieu}\ena

Here, overdots denote derivatives with respect to conformal time
(e.g. $\dot{x} = a\frac{dx}{dt}$), and in the second
equation we have replaced $U_{eff} = \frac{\ddot{a}}{a}$, which
acts as a time-dependent \emph{effective
potential} \cite{grishchuk93a}. We recognize Equation
\ref{e:gwb_mathieu} as a version of Mathieu's equation for the
vertically driven pendulum, Equation \ref{e:mathieu}, once the following
substitution is made in Equation \ref{e:mathieu}:

$$-\Omega^2 \cos{2 \pi f t} \equiv U_{eff}(a)$$ In contrast to the vertically driven pendulum, variation
of the pump parameter $U_{eff}$ does not lead to runaway growth.
Rather, the time-varying effective potential amplifies
long-wavelength oscillations relative to short wavelength
oscillations.

For an isotropic, homogenous universe consisting of a fluid with
pressure $p$ and density $\rho$ we can express

\bea
U_{eff}(a) &=& \frac{d}{dt}\Big(a\frac{da}{dt}\Big)\nonumber\\
&=& \Big(\frac{da}{dt}\Big)^2 + a\frac{d^2a}{dt^2} = a^2 H^2 + a\frac{d^2a}{dt^2}\nonumber\\
&=& a^2 \Big[\frac{8\pi G\rho}{3} - \frac{4\pi G}{3} (\rho +
3p/c^2)\Big] \label{e:ueff}\ena

Here, we have employed the definition of the Hubble parameter
$H(a)\equiv \frac{1}{a} \frac{da}{dt} = \sqrt{8\pi G\rho(a)/3}$.
For convenience, the equation-of-state relating pressure, $p$, and
density, $\rho$, is taken as $p = \gamma\rho c^2$ (where $\gamma$
is a scalar that depends on the cosmological epoch under
consideration). As the universe expands, it dilutes: $\rho =
\rho_*(\frac{a}{a_*})^{-3(1+\gamma)}$. This equation is valid at
any epoch, or correspondingly, for any value of $a$. When we
consider a \emph{specific} epoch we label it $a_*$. From Equation
\ref{e:ueff} we obtain

$$U_{eff}(a) = \frac{8\pi G\rho a^2}{3}\Big[1-\frac{(1+3\gamma)}{2}\Big]
= \frac{4\pi G \rho_*(1-3\gamma)
a^2_*}{3}\Big(\frac{a}{a_*}\Big)^{2-3(1+\gamma)}$$ which can be
written as

\beq U_{eff}(a)=\frac{4\pi
G\rho_*(1-3\gamma)a^2_*}{3}\Big(\frac{a}{a_*}\Big)^{-(1+3\gamma)}\label{e:ueffgamma}\eeq

The evolution of the effective potential depends crucially on the
cosmological epoch, via the relationship between density and
pressure. For example, during radiation domination, $\gamma = 1/3$
and the effective potential $U_{eff}(a) = 0$. Recalling Equation
\ref{e:gwbpara}, when either $U_{eff}(a) = 0$ or $k^2\gg U_{eff}$,
the solutions for $\nu$, the gravitational wave amplitude, are
simple plane-waves. These purely oscillatory solutions prevail
\emph{whenever} $k\gg U_{eff}$, but especially during radiation
domination when the dilution of the GWB is identical to that of
\emph{any} radiation background, such as the CMB.

For future reference, using the definition of the Hubble parameter
$H(a)$ in terms of density $\rho(a)$, we can write \beq
U_{eff}=\frac{(1-3\gamma)k^2_*}{2}\Big(\frac{a}{a_*}\Big)^{-(1+3\gamma)}
\label{e:ueffhubble}\eeq where $k_{*} \equiv a_* H(a_*)$.

\begin{figure}
\centering
\includegraphics[height=7cm,angle=0]{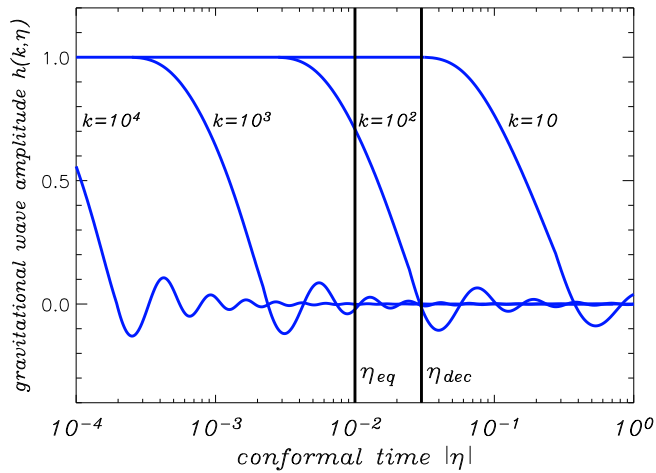}
\caption{Gravitational wave amplitude as a function of conformal
time for four different wavenumbers. Gravitational waves are
constant, and equal in amplitude independent of wavelength, before
entering the horizon when $k\eta \sim 1$, leading to decay.
Equivalently, waves with $k \ll U_{eff}(a)$ (long wavelength
waves) experience the effective potential, forestalling their
decay. Short wavelength waves never interact with the potential,
continuously decaying adiabatically instead. Long wavelength modes
essentially ``tunnel" through the barrier with no diminution of
their amplitude until after radiation-matter equality. Figure
credit: Nathan Miller.} \label{f:h}\end{figure}

Using Equations \ref{e:h_eq_nu_div_a} and \ref{e:ueffgamma}, we can
solve Equation \ref{e:gwbpara} for the gravitational wave amplitude

\beq h =\sqrt{8\pi G} \, \frac{a_*}{a}e^{[ik(\eta-\eta_*)]}
\label{e:h}\eeq  Gravitational waves ``enter" the horizon when
$k\simeq 1/\eta$, and as the universe Hubble-expands, they are
damped by the adiabatic factor $1/a$, just like radiation. Example
solutions of Equation \ref{e:h} are shown in Figure \ref{f:h}.

\subsection{The effective potential}

We have shown that the equation-of-state parameter $\gamma$
determines the effective potential, subsequently determining the
evolution of the GWB, at least during radiation domination. In
this subsection we examine solutions in the other important
cosmological epochs. We will see that specifying $\gamma$ versus
cosmological scale not only allows us to predict the advantages of
indirect detection of the GWB (using B-modes) over direct
detection methods, but also to \emph{optimize} experimental CMB
polarization surveys themselves.

In the following, the subscripts ``-" and ``+" will denote
quantities before and after inflation ends, respectively. If
$a_{end}$ denotes the scale factor \emph{at} the end of inflation,
this means that for $a < a_{end}, \gamma < -\frac{1}{3}$. After
inflation ends $a>a_{end}$ and $\gamma_+
>-\frac{1}{3}$. For example, $\gamma_+ = 1/3$ corresponds to the
equation-of-state for radiation; that is, when inflation ends
the universe is radiation dominated.

More generally we can say that at the moment when $a=a_{end}$,
accelerated expansion ($d^2a/dt^2 > 0$) changes to decelerating
expansion ($d^2a/dt^2 < 0$), and during inflation the cosmological
horizon \emph{decreases} \cite{liddlelyth2000}. Hence, from Equation
\ref{e:ueffhubble}, for $a < a_{end}$,

\beq
U_{eff}=\frac{(1-3\gamma_-)k^2_*}{2}\Big(\frac{a}{a_*}\Big)^{|1+3\gamma_-|}\label{e:ueffgrow}\eeq which increases with $a$, and when $a > a_{end}$

\beq
U_{eff}=\frac{(1-3\gamma_+)k^2_*}{2}\Big(\frac{a}{a_*}\Big)^{-|1+3\gamma_+|}
\label{e:uffshrink}\eeq which decreases with $a$. Since the transition from $\gamma_-$ to
$\gamma_+$ occurs quasi-instantaneously, the effective potential is
discontinuous;  $\gamma_+>\gamma_-$. For more accurate results we
should properly treat the reheating phase, when the inflaton is
converted into particles and radiation. Interestingly, parametric
resonance techniques can also be used to describe reheating
\cite{kofman94}.

Now it should be clear why $\frac{\ddot{a}}{a}$ was called an
\emph{effective potential}. Since $U_{eff}$ is maximized when
inflation ends, there are two epochs, $a_- < a_{end}$ and $a_+ =
a_{end}$, where $k^2 = U_{eff} < U_{max}$. Thus there are two
wavelength regimes that determine the form of solutions to Equation
\ref{e:gwbpara}. High-frequency waves with $k/k_{H} \gg \eta_{eq}$
enter the horizon well before matter-radiation equality, then
decay as the universe expands as in Equation \ref{e:h}
\cite{dodelson02,deepak06}.

Long wavelength waves, on the other hand, satisfy \beq \ddot{\nu}
- U_{eff}(a)\nu = 0 \label{e:long}\eeq To solve Equation \ref{e:long}
we must first determine the behavior of the effective potential in
a form that is valid during any cosmological epoch.

\subsection{Timing is everything: the cosmic chronology}

To analyze the impact of the effective potential on long
wavelength gravitational waves we must solve for $U_{eff} \sim
\ddot{a}/a$, recalling that the derivatives are with respect to
conformal time, $\eta$. We therefore require the relationship
between the scale factor and conformal time. For reference, we
recall that during radiation domination $U_{eff}=0$ for all
wavelengths.

\subsubsection{Evolution of the effective potential}
\begin{figure}
\centering
\includegraphics[height=9cm,angle=0]{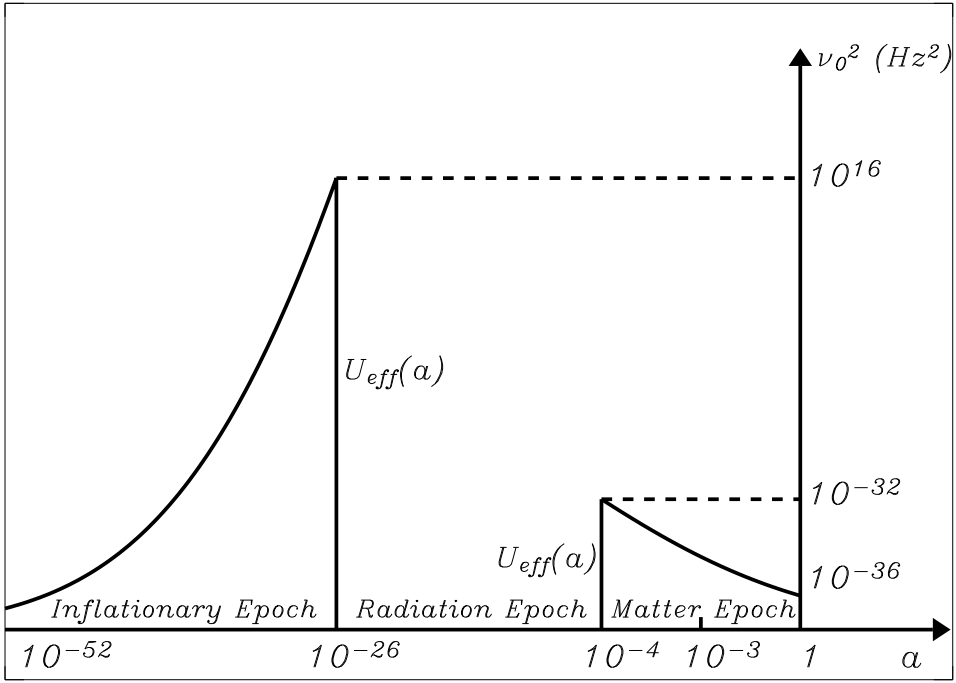}
\caption{The parametric amplification of primordial gravitational
waves is governed by the evolution of the effective potential as a
function of cosmic scale factor, $a$. This figure, adapted from
\cite{grishchuk93a}, shows the effective potential in three
important cosmological epochs. Here the wave's frequency ($\nu_0 =
c k$) is expressed in present-day units, when $a=1$. Long
wavelength waves ($k^2<U_{eff}$) ``tunnel" through the potential
and remain constant until inflation ends. Short wavelength waves
never experience the potential and instead decay adiabatically (as
$1/a$). During radiation domination, the effective potential
vanishes and all waves inside the horizon decay. Finally, any
waves that survive until matter-radiation equality imprint the CMB
sky prior decoupling. Therefore, these waves are comparable to, or
larger than, the horizon at decoupling, subtending an angle of
$\simeq 2\arcdeg$ on the sky today. Figure credit: Nathan Miller.}
\label{f:para_amp}
\end{figure}

During the epoch of matter domination, it is convenient to
parameterize the evolution of the scale factor versus time as $a
\sim t^\alpha$ with $\alpha = 2/3$. Using the definition of
conformal time, we have $d\eta = dt/a$, implying that
$a=\eta^{\frac{\alpha}{1-\alpha}}$ or $a(\eta) = \eta^2$ during
matter domination. Using this, we find that the effective
potential during matter domination \emph{decays} as $U_{eff}(a)
\propto 1/a$, as described quantitatively in Equation \ref{e:uffshrink}
and displayed in Figure \ref{f:para_amp}.

Finally we must calculate the effective potential during
inflation, when the scale factor grows as $a = e^{Ht}$, implying
$d\eta = e^{-Ht} dt$. Integrating, we find $\eta=(1-e^{-Ht})/H$ or
that \beq a \simeq \frac{1}{H|\eta|}\label{e:heta}\eeq during
inflation, leading to \beq U_{eff}(a) = 2 H^2
a^2\label{e:ueffha}\eeq That is, the effective potential
\emph{grows} quadratically.

\subsubsection{Evolution of the GWB}

With the effective potential expressed in terms of conformal time
we can easily solve the gravitational wave equation. During
\emph{inflation}, $a\sim 1/\eta$ so $U_{eff} \propto 2/\eta^2$ (as
anticipated from Equation \ref{e:ueffgrow}). For long wavelength waves,
the parametric equation is

\beq \ddot{\nu} - \frac{2}{\eta^2} \nu = 0 \label{e:long2}\eeq

Solutions to Equation \ref{e:long2} are easily verified to be $\nu \sim
1/\eta$, or $\nu(a) \sim a$. Since $h(a) \equiv \sqrt{8 \pi G}
\nu/a$ (Equation \ref{e:h_eq_nu_div_a}) we see that the gravitational
wave amplitude during the inflationary epoch is constant for waves
with wavelengths smaller than $U_{eff}(a<a_{end})$, or
equivalently

\beq h(k,\eta)=\frac{\sqrt{8\pi G}}{\eta a}\label{e:hinfl} \eeq

During \emph{radiation domination}, $U_{eff} = 0$ and the
gravitational waves are oscillatory functions that redshift and
dilute adiabatically as

$$h = h_{-}\frac{a_-}{a}e^{[ik(\eta-\eta_-)]}$$

During \emph{ matter domination}, although the effective potential
has the same dependence on $\eta$ as it does during inflation, the
scale factor depends on conformal time in a different way. This
leads to a \emph{different} gravitational wave solution. During
matter domination, $\eta = \sqrt{a}$ and so $h = \nu/a \sim
1/a^{3/2}$, implying that gravitational waves decay even when
their wavelengths are long ($k<U_{eff}(a_{eq}))$. Once again,
however, longer wavelengths are preferentially preserved relative
to short wavelength modes.

Thus we have the following cosmic chronology. During inflation,
long wavelength gravitational waves are ``frozen" outside the
horizon, or equivalently, are spared from adiabatic decay as they
tunnel through the effective potential, which grows quadratically
as the universe inflates. At the beginning of radiation domination
(end of inflation; $a_{end}$), the effective potential drops to
zero. All waves that are within the horizon decay adiabatically.
The largest waves persist until decoupling, when the universe is
matter dominated and the potential again becomes critically
important. The primordial scale-invariant distribution of waves is
transformed twice; first during inflation, and later during matter
domination. In both cases, longer wavelength perturbations are
overpopulated with respect to their short wavelength counterparts.



\subsection{Quantum gravitational wave effects}

We can make an analogy between the GWB and the laser here by
rewriting the GWB parametric equation (Equation \ref{e:gwbpara}) as
$\ddot{\nu} +(k^2-U_{eff}(a))\nu = \ddot{\nu} +k_{eff}^2\nu =0.$
For $k_{eff}^2>0$ the solutions are constant oscillatory
functions, while for $k_{eff}^2<0$ we have decaying solutions.
Similar derivations \cite{dodelson02,deepak06}, using creation and
annihilation operators, are particularly useful for exploring the
connections between the quantum properties of the GWB and the
laser (which can also be derived using creation and annihilation
operators). This derivation is possible because the gravitational
waves, or gravitons, are bosons, as are the laser's photons. In
both cases, the pump need not be periodic.

However, the analogy cannot be taken much further. For the
prototypical three-level laser, coherent amplification is obtained
via stimulated emission from a metastable state. The metastable
state's occupation number is inverted with respect to the ground
state. For the GWB there is no such population inversion, nor
stimulated emission, and thus the GWB is an incoherent, stochastic
radiation background, not unlike the CMB.

We are interested not only in the properties of a \emph{single}
gravitational wave, but also in the behavior of a stochastic
ensemble of waves, tracing its origin to ``quantum noise" in the
inflaton. To quantify the preferential population of long
wavelength gravitational waves, we calculate the initial number of
gravitons, or occupation number, for given conformal wave number
$k$ when $a\ll a_{end}$, denoted as $N_{in}$. We construct the
correlation function of this stochastic background by considering
the energy of a collection of $N$ oscillators, each with energy
$E=\hbar\omega(n+\frac{1}{2})$ in state $n$. Using terminology
familiar from laser physics, the number of
gravitons \cite{Polnarev2006} in a volume $V(a)$ is the
(renormalized) energy divided by the product of the frequency and
the reduced Planck constant, $\hbar$:

\beq N_{in}=\frac{c^4
V(a)}{32G\pi\hbar\omega(a)}\Big\langle\Big(\frac{dh_\alpha^\beta}{dt}\Big)
\Big(\frac{dh_\beta^{\alpha^{*}}}{dt}\Big)\Big\rangle\label{e:nin}
\eeq

In Equation \ref{e:nin}, $*$ indicates complex conjugation, the angled
brackets $\langle \ldots \rangle$ denote averages over
polarization, propagation angles, time, and volume. The average
results in a constant factor that is subsequently incorporated
along with the physical constants into a constant $\kappa$.
Finally, we define $\omega \equiv k/a$, thus simplifying the time
derivatives: $\frac{dh^\alpha_\beta}{dt}=\omega h^\alpha_\beta$
and $V(a) = V_0 a^3$.

As result we have

\bea N_{in} &\approx& \kappa
\Big(\frac{h_-a_-}{a}\Big)^2\Big(\frac{k}{a}\Big)^2
\Big(\frac{k}{a}\Big)^{-1} V_0 a^3\\
&\approx& \kappa h^2_-a^2_-kV_0 \ena which does not depend on $a$. Similarly, the \emph{final} mode
occupation number at the end of inflation, $N_{out}$, when $a \gg
a_{end}$ and $k^2 \gg U_{eff}$ is

$$N_{out} \approx \kappa h^2_+a^2_+kV_0$$ which is also independent of time and $a$.
For $a_-< a <a_+$, when $k\ll U_{eff}$ (long wavelength
gravitational waves), the solution for the gravitational wave
amplitude is $\nu\approx a$. For a scale invariant spectrum $h$ is
constant, meaning that gravitons are created with the same
amplitude ($h_-\approx h_+$). This leads to the amplification
factor:

\beq A=\frac{N_{out}}{N_{in}} \approx \Big(\frac{a_+}{a_-}\Big)^2
\label{e:amp} \eeq which is larger than unity if $k^2<U_{max}$.
Therefore, long wavelength gravitational waves are amplified with
respect to their short wavelength counterparts (which experience
the potential for a much shorter time). This is evident from Figure
\ref{f:para_amp}. Intermediate wavelength waves ($k^2 \sim
U_{max}$) do not experience the effective potential until much
later than their long wavelength counterparts, and very short
wavelength waves, with $k^2 > U_{max}$, do not enter the potential
at all.

As such, if all waves start with the same initial amplitude
(Harrison-Zeldovich spectrum), short wavelength waves will decay
by a quadratic factor relative to longer waves. We note that the
change in scale factor during inflation can be enormous; expansion
by factors of $10^{30}$ are common for models that produce
sufficient inflation to make spacetime flat. In the context of
the GWB, only the \emph{difference} in the expansion of the
universe between the horizon-entry time for short wavelength modes
compared to that for long wavelength modes is relevant.

\section{Observational consequences}

To illustrate the significance of parametric resonance, it is
useful to consider the following generic inflation scenario where
the equation-of-state parameter changes from $\gamma_-=-1$ to
$\gamma_+=1/3$; that is, radiation domination immediately
follows inflation. Using Equation \ref{e:ueffhubble}, in such a
scenario a wave with $k_{end} \simeq a_{end}H_{end}$ enters the
effective potential just as inflation ends. As such it is not
amplified. A long wavelength wave enters the potential earlier
when $a_- \simeq H/k$. We are free to define the scale factor at
the end of inflation as $a_{end}=1$. Thus from Equation \ref{e:amp},
with $a_+=a_{end}=1$, the amplification scales with wavevector as
$A\sim \Big(\frac{H^2}{k^2}\Big)$, assuming $H$ is constant during
inflation. The implication of this equation is clear: if the
primordial gravitational wave spectrum is scale invariant ($h_-
\simeq h_+$), then post-inflation it is transformed into a
strongly wavelength-dependent spectrum.

Not all of these amplified waves survive to imprint the CMB with
B-mode polarization. Waves that are larger than the present-day
horizon are frozen and are not amplified. Additionally, all
sub-horizon waves decay by the adiabatic factor
$\frac{a_{rad}}{a_{eq}}$ from the onset of radiation domination at
reheating to the epoch of matter-radiation equality.

\subsection{Tensor power spectrum: amplitude and angular structure} More generally, we are interested in the variance of
$h(k,\eta)$ as a function of wave number, also called the
\emph{tensor power spectrum}, defined as $$ P_t(k) \equiv
\frac{|h(k,\eta)|^2}{k^3}$$ From Equation \ref{e:hinfl} we recall that
$ h(k,\eta)=\frac{\sqrt{8\pi G}}{\eta a}$.  Since $\eta \simeq
1/aH$ (Equation \ref{e:heta}), we find

\beq P_t(k) = \frac{8 \pi G H^2}{k^3} \label{e:psubt} \eeq

Thus, measuring the tensor power spectrum probes the Hubble
constant during inflation and is proportional to the energy
density of the inflaton (since $H^2\propto \rho$). The amplitude
of the tensor power spectrum is directly revealed by the amplitude
of the CMB curl-mode polarization power spectrum. If inflation
occurs at energy scales comparable to $E_{GUT}$, CMB B-mode
polarization allows us to test physics at scales one trillion
times higher than the highest energy produced in terrestrial
particle accelerators!

We have shown how the tensor power spectrum's amplitude depends on
the energy scale of inflation. Now our aim is to qualitatively
determine the angular correlation properties, or shape, of the
tensor power spectrum $P_t(k)$. Doing so allows us to optimize
observations of the GWB, for as we will see, $P_t(k)$ has a
well-defined maximum as a function of wavenumber, or equivalently,
angle subtended on the sky.

In the context of the driven undamped pendulum, the \emph{ansatz}
solutions, Equation \ref{e:ansatz}, can grow exponentially or remain
constant. The \emph{long} wavelength gravitational waves
amplitudes are also stable when they are larger than the
cosmological horizon. Short wavelength gravitational waves decay
adiabatically because they are always within the horizon during
inflation. In the context of parametric amplification, long
wavelength waves ($k^2<U_{eff}$) are said to be ``super-horizon"
or non-adiabatic.

Since the extremely high-frequency waves are always above the
potential barrier, they continuously decay and leave no observable
signature. The longest waves enter the horizon near decoupling
when the effective potential is decaying. Waves that experience
the matter-dominated effective potential close to decoupling are
amplified the most, leading to a well-defined peak in the tensor
power spectrum on scales comparable to the horizon at decoupling.
This scale subtends an angle of $\simeq 1\arcdeg$ to $2\arcdeg$ on
the CMB sky today, and therefore this is the characteristic scale
of the tensor and B-mode angular correlation function.

\subsection{Direct versus indirect detection of the primordial GWB}

It is perhaps instructive to ask whether the primordial GWB could
be directly detected \emph{today}. While scalar, or mass-energy,
perturbations are amplified by gravitational condensation, the
tensor GWB is not. Directly detecting the GWB \emph{today}
(redshift $z = 0$) by, for example, LIGO, would be extremely
difficult since, like the primordial photon background (the CMB),
the energy density of the primordial GWB dilutes (redshifts) by
the fourth power of the scale factor as the universe expands.
However, the GWB imprints curl-mode polarization on the CMB at the
surface of last scattering (at $z \sim 1,100$ or 380,000 years
after the Big Bang). Therefore, the energy density of the GWB at
last scattering was more than \emph{one trillion times}
($1,100^4$) larger than it is now.

Ultimately, direct detection experiments such as ESA's
\textit{LISA}, NASA's \emph{Big Bang Observer}, or Japan's
\emph{DECIGO} experiment will provide further tests of the
inflationary model, such as measuring the GWB power spectrum at
wavelengths approximately twenty orders of magnitude smaller than
those probed by CMB polarization \cite{chongefstat06,
smithkam06,smithpierpaolikam06}. As we have seen, the current
spectral density of these short wavelength waves is extremely
small, and these experiments are fraught with contamination from
``local" (i.e. non-cosmological) sources. However, given a
detection of the primordial GWB at the surface of last scattering
using CMB polarization, these direct detection campaigns will
measure the fine details of the inflaton potential, making them at
least well justified, if not mandated.

\subsection{Indirect detection of the GWB: optimizing CMB polarization observations}
Parametric amplification had important ramifications for the
design of the first experiment dedicated to measuring the
GWB---BICEP
(Background Imaging of Cosmic Extragalactic Polarization).
While the amplitude of the GWB is
unknown, parametric amplification allows the structure of the
B-mode's angular correlation function to be accurately calculated
given only modest assumptions about the cosmic equation-of-state
and evolution of the scale factor. This allowed the BICEP team to
optimize the angular resolution of our search for the B-mode
signature; motivating both BICEP's optical design as well as its
survey design (required sky coverage).

BICEP probes the inflationary GWB primarily at
large angular scales corresponding to the largest
wavelength waves that entered the horizon near decoupling
($\simeq 2\arcdeg$). Since the tensor angular correlation function
peaks on these scales, to obtain statistical confidence in our
measurement it is only necessary for BICEP to probe a small
fraction of the sky ($\simeq 3\%$), rather than diluting our
observing time over the entire sky. This allows us to target the
cleanest regions of the sky---those with minimal contamination
from galactic dust or synchrotron radiation, both of which are
known to be polarized.


\subsection{Detectability of the CMB B-mode polarization}

Nearly thirty years passed between the discovery \cite{Penzias65} of
the CMB by Robert Wilson and Arno
Penzias (Charles Townes' Ph.D. student) and the first detection of temperature
anisotropy \cite{smoot92} by COBE at the ten parts-per-million
level. If inflation occurred at the GUT-scale, it would produce
curl-mode polarization at the ten parts-per-\emph{billion} level.

Thanks to the innovative technologies produced by our
collaboration, detection of this signal is conceivable.
BICEP---the ground-based polarimeter we built---will
achieve higher
sensitivity to the GWB than either NASA's WMAP (Wilkinson Microwave
Anisotropy Probe) or the Planck satellite.

BICEP is both a pioneering experiment and a long-range, two-phase
campaign designed to mine the CMB sky using innovative technology.
BICEP is an attempt to probe even farther back than the last
scattering surface: to the very beginning of the universe, the
inflationary epoch. Detecting the GWB requires ultrasensitive
technology, only recently invented. An understanding of parametric
amplification allows us to precisely estimate the spatial power
spectrum imprinted on the CMB's polarization by the primordial
GWB. This leads to a very general optimization of experimental
campaigns \cite{jaffe2000}. To detect the GWB's polarization
imprint, only modest angular resolution is required---corresponding
to a small refractor. This small refractor can probe
the inflationary B-mode polarization nearly as well as a
reflecting telescope twenty-times larger in diameter!

As shown in the following section, BICEP's small size has several
important ancillary benefits---most notably that its smaller
aperture results in a much higher-fidelity optical system, one
with no obscuration or secondary mirror to induce spurious
polarization. Additionally, the refractor is easy to shield from
stray light---which is particularly important when probing signals one
billion times smaller than the background.

\section{Experimental quantum cosmology: the BICEP project}
BICEP \cite{keating2003, yoon06} is a bold first step toward
revealing the GWB. Ultimately, only a small telescope like BICEP
can be cooled entirely to nearly the temperature of the CMB itself---a
condition not previously achieved, even in space. BICEP's
elegant design (see Figure \ref{f:BICEP}) has proven extremely attractive for proposed future
experiments \cite{sampan, montroy2006SPIE}, which have receiver
concepts closely resembling BICEP.

BICEP is the first experiment to directly probe for the primordial
GWB. Even Planck, when it is launched in 2008, will not be as
sensitive to the GWB signal as BICEP (which will already have
completed first-phase observations). A comparison of BICEP's
capability to detect the inflationary GWB with that of WMAP and
Planck is demonstrated in Hivon \& Kamionkowski \cite{hivon02}.
They show that BICEP will achieve higher sensitivity to the GWB
than these spaceborne experiments. BICEP's sensitivity results
from advances in detector technology and from the ability to
target only the cleanest regions of the microwave sky, rather than
spreading out limited integration time over the full sky.

\subsection{Optics}
\begin{figure}
\centering
\includegraphics[height=14cm,angle=0]{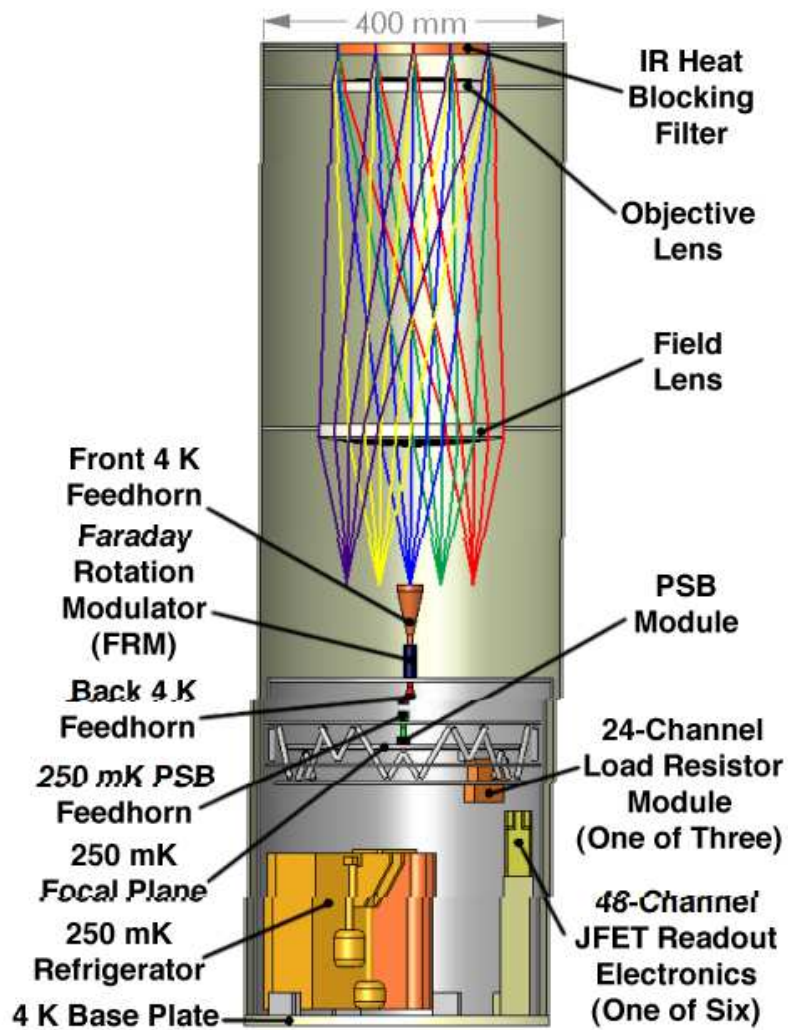}
\caption{A cross-sectional view of the BICEP receiver, which
comprises a refracting telescope and 49 polarization-sensitive
bolometers. The optics and focal plane are housed within a
cryostat, which is placed on a three-axis mount. Figure credit:
Thomas Renbarger.} \label{f:BICEP}
\end{figure}

As outlined above, only modest angular resolution
is required to detect the GWB's
polarization signature. BICEP was designed to map $\sim 3\%$ of
the sky with $0.9 \arcdeg$ resolution (at 100 GHz), $0.7\arcdeg$
resolution (at 150 GHz), and $0.5\arcdeg$ resolution (at 220 GHz).
Unlike Planck or WMAP, BICEP was designed specifically for CMB
polarimetry, able to modulate the polarization signal independent
from the temperature signal with high fidelity. BICEP achieves
this fidelity by virtue of an elegant optical design: a 4 kelvin
refractor (Figure \ref{f:BICEP}). Millimeter-wave radiation enters
the instrument through a 30 cm diameter vacuum window and passes
through heat-blocking filters cooled to 4 K by liquid helium. Cold
refractive optics produce diffraction-limited resolution over the
entire $18 \arcdeg$ field-of-view.

\subsection{Detector system}
Each one of BICEP's 49 pixels comprises a complete polarimeter
(optics, polarization modulator, analyzer, and detector). Each
pixel uses three corrugated feedhorns feeding a
polarization-sensitive bolometer pair (PSB), which simultaneously
analyzes and detects linearly polarized light. The PSB's ingenious
design is further described in \cite{jones03}. In addition to their use in
{BICEP}, PSBs have been successfully used in the \emph{BOOMERANG}
\cite{montroy06} and \emph{QUAD} experiments to measure the E-mode
polarization signal, and they will be used on Planck.

BICEP's PSBs use two absorbing grids, each coupled to a single
linear polarization state. The (temperature-dependent) resistance
of a semiconducting thermistor (neutron transmutation doped
germanium; \emph{NTD-Ge}), located at the edge of the absorber,
detects CMB photons via a resistance change. BICEP's PSBs have
astounding sensitivity: in one second, each PSB can detect
temperature fluctuations as small as
$\approx 450~\mu \rm{K}$.

\subsection{Polarization modulation}

The faintness of the GWB polarization signal demands exquisite
control of instrumental offsets. There are two ways to mitigate
offsets: (1) minimize the offset and (2) modulate the signal before
detection (Dicke switch) faster than the offset fluctuates. BICEP
does both. A bridge-circuit differences the two PSBs within a
single feed, producing a (first) difference signal that is null
for an unpolarized input. This minimizes the offset. For six of
the 49 spatial pixels, the polarized signal input is rapidly
modulated (second difference) by Faraday Rotation Modulators (FRM)
that rotate the plane of linear polarization of the incoming
radiation. The FRMs make use of the Faraday Effect in a magnetized
dielectric. Polarization modulation allows the polarized component
of the CMB to be varied \emph{independently} of the temperature
signal, allowing the response of the telescope to remain fixed
with respect to the (cold) sky and (warm) ground. This two-level
differencing scheme allows for two levels of phase-sensitive
detection, allowing optical systematic effects, associated with
the telescope's antenna response pattern (leaking the much-larger
CMB temperature signal to spurious CMB polarization), to be
distinguished from true CMB polarization.

The FRMs represent a significant advance in the technology of CMB
polarization modulation. Early CMB polarimeters (including Penzias
and Wilson's, which \emph{was} polarization-sensitive) used
rotation of the entire telescope to modulate CMB polarization.
These experiments \cite{lubin79,nanos79,keating2003b} rotated
hundreds or thousands of kilograms, were susceptible to vibration
induced microphonic noise, and were limited mechanically to
modulation rates $<0.1$ Hz.

The next polarization modulation innovation was a birefringent
half-waveplate: a single crystal of anisotropic dielectric
(typically quartz or sapphire) that phase-delays one of the two
linear polarizations \cite{caderni78,philhour02,johnson2003}.
While the fragile $\sim 1$ kg, cryogenically-cooled crystal
\emph{can} be rotated at $\sim 1$ Hz with lower-vibration than
rotation of the entire telescope, such a mechanism is prone to
failure since bearing operation is a severe challenge at cryogenic
temperatures. And since bolometers are sensitive to power
dissipation at the $10^{-17} $W level, even minute mechanical
vibrations produced by the bearings are intolerable.

\begin{figure}
\centering
\includegraphics[height=12cm,angle=0]{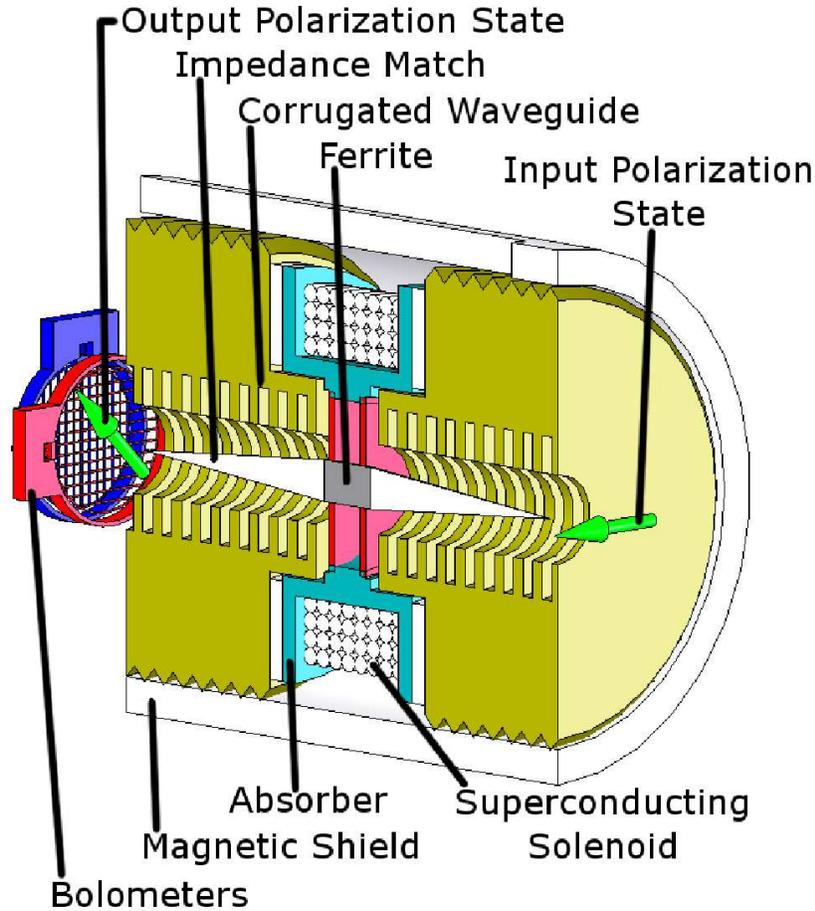}
\caption{A cross-sectional view of a Faraday Rotation Modulator
(FRM) pixel. Polarized light enters from the right, is rotated by
$\pm 45\arcdeg$ and then analyzed (decomposed into orthogonal
polarization components, which are detected individually by the
polarization-sensitive bolometers, or PSBs). For each of BICEP's
pixels, the PSBs are contained within a corrugated feedhorn,
cooled to 0.25 K, located approximately 20 cm from the FRM, which
is placed in a corrugated waveguide at the interface between two
corrugated feedhorns, placed back-to-back and cooled to 4 K. The
(schematic) location of the PSBs in this figure serves to
illustrate the coordinate system used as the polarization basis.
Figure credit: Thomas Renbarger.} \label{f:frm}
\end{figure}

Faraday Rotation Modulators\footnote{US Patents Pending: ``Wide
Bandwidth Polarization Modulator, Switch, and Variable
Attenuator," US Patent and Trademark Office, Serial Number:
60/689,740 (2005).},
shown in Figure
\ref{f:frm}, require only ``rotating" electrons (the generation of
a solenoidal magnetic field in a magnetized dielectric) to effect
polarization rotation. Therefore, FRMs reduce the rotating mass
that provides modulation by 30 orders of magnitude! Furthermore,
these devices are capable of rotating polarized millimeter wave
radiation at rates up to 10 kHz---faster than any conceivable
time-varying temperature- or electronic-gain fluctuation. A
superconducting NbTi solenoid wound around the waveguide provides
the magnetic field that drives the ferrite into saturation,
alternately parallel and anti-parallel to the propagation
direction of the incoming radiation. The FRM rotates the CMB
polarization vectors by $\pm±45\arcdeg$ at 1 Hz, well above
1/f-fluctuation timescales (caused by, for example, temperature
variations). The bolometer signals from the PSBs are detected
using lock-in amplification.

\subsection{Observations of galactic polarization using Faraday Rotation Modulators}

Initial observations of the galactic plane were obtained during
the austral winter of 2006. Several hundred hours of data were
taken with the FRMs biased with a 1 Hz square wave. This
modulation waveform effected $\pm 45\arcdeg$ of polarization angle
rotation. Other modulation waveforms can provide more or less
rotation, as desired.

\begin{figure}
\vspace{1cm}
\includegraphics[height=15.cm,angle=0]{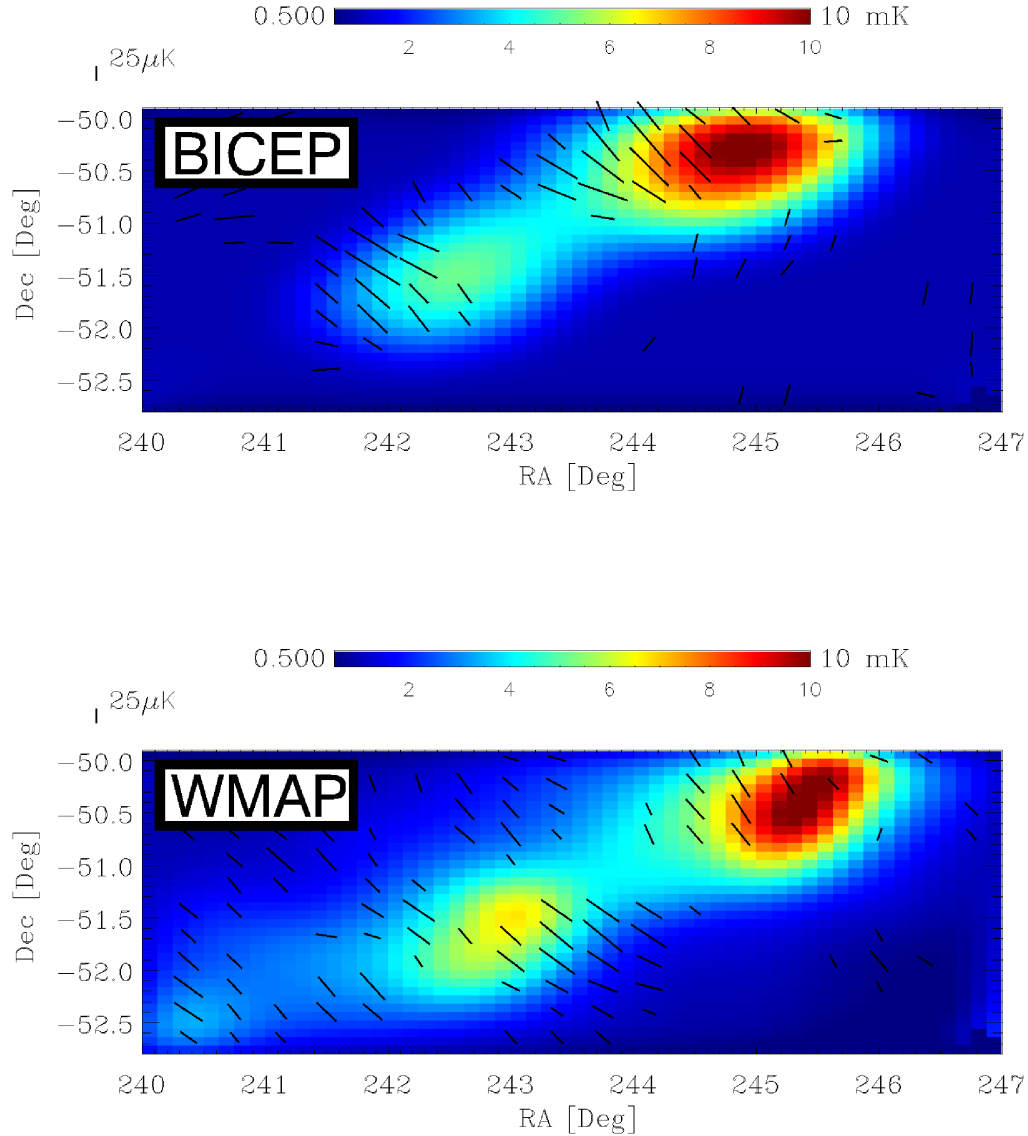}
\caption{Map of a portion of the galactic plane made using one of
BICEP's six Faraday Rotation Modulator (FRM) pixels (top)
operating at 100 GHz, compared to WMAP's observations of the same
region (bottom). Microwave radiation is polarized by dust grains
preferentially aligned by the galaxy's magnetic field. The FRM
pixels modulate only the polarized component of the emission and
can fully characterize linear polarization \emph{without} rotation
of the telescope. The short lines indicate the magnitude and
orientation of the plane of polarization, and the gray scale
indicates the temperature scale. For comparison, a scale bar
representing
$25~\mu$K linear polarization is shown. The galactic plane extends
approximately from the lower left to the upper right of each map.
Both maps show significant linear polarization orthogonal to the
galactic plane, which is expected as the galaxy's magnetic field
is oriented parallel to the plane \cite{page2006}. Similar maps
were produced for BICEP's 150 GHz FRM pixels. We note that the
BICEP FRM data were acquired over the course of a \emph{week} of
observations, whereas the WMAP data was obtained over three years.
Figure credit: Evan Bierman.\label{f:frm_maps} }
\end{figure}

To validate the FRM technology we targeted several bright regions
of the galactic plane. Results from some of these observations are
displayed in Figure \ref{f:frm_maps}, where we also show the same
region as imaged by WMAP \cite{page2006}. The agreement is
impressive. Because BICEP's bolometers simultaneously measure
polarization and temperature anisotropy, we use WMAP's temperature
maps as a calibration source for BICEP.

\subsection{The South Polar Observatory}

Essential to achieving maximum sensitivity to CMB B-modes is the
long integration time afforded by the South Pole site---arguably
the premiere, long-duration, low-background (both natural and
manmade) Earth-based site. To exploit this location, BICEP is
highly efficient, having been designed for robustness and
quasi-autonomous operation, while consuming a minimum of liquid
cryogens (a precious commodity at the South Pole). BICEP's
toroidal cryogen tanks house the instrument in a thermally uniform
4 K environment. The 49 polarimeter pixels, optics, and sub-Kelvin
refrigerator are removable for easy instrument servicing. In
December 2005, BICEP was installed at the US Amundsen-Scott
South Pole Station's Dark Sector Laboratory, operated by
the National Science Foundation. The Dark Sector Laboratory will
house BICEP for three austral winter observing seasons and then
house a future upgraded version, called BICEP-II, maximally
leveraging the polar infrastructure investment for years to come.

More detail about the design of BICEP can be found in Keating
\emph{et al.} \cite{keating2003}. Preliminary data, maps, and
additional technical information from BICEP's first observing
season can be found in Yoon \emph{et al.} \cite{yoon06}.

\section{Future probes of the past: BICEP-II, an advanced CMB polarimeter array}

Following BICEP's initial phase, we will deploy an advanced
high-density array to our South Pole observatory, eventually
yielding  ten-times better sensitivity than the first phase of
BICEP. The GUT-scale is already nearly within the reach of current
experiments \cite{keating2006}, such as BICEP, and nearly all
detectable inflationary models will ultimately be testable with
BICEP-II's technology.

BICEP's NTD-Ge semiconducting detectors are background-limited,
meaning the sensor's intrinsic noise is sub-dominant compared to
the photon noise from atmospheric emission. In this regime,
increasing the signal-to-noise ratio of the experiment can  be
accomplished only by adding detectors. After BICEP's three-year
campaign concludes, our team will increase the number of detectors
in BICEP's focal plane by a factor of more than five. This will
produce a CCD-array-like focal plane which we call ``BICEP-II."

BICEP-II will use the same optical design, observatory, and
observing strategy as BICEP, but will be upgraded to an advanced
detector-array of superconducting transition-edge sensors (TES).
The TES \cite{bockbolo03} replaces the NTD-Ge
semiconductor thermistor bonded to each absorbing grid in the
current PSB design with a superconductor operated near its
normal-to-superconducting transition temperature. CMB photons heat
up the superconductor, causing its resistance to change enormously,
which makes it an ideal sensor.

BICEP requires three electroformed feedhorn antennas per pixel
both to receive power from the sky and to couple millimeter wave
radiation to the bolometers, see Figure \ref{f:BICEP}. The
electroforming process is costly and time-consuming, and the
feedhorn's dimensions fundamentally limit the packing efficiency
of the focal plane. The TES methodology has a significant
logistical advantage: the superconductor and associated components
are all fabricated photolithographically, resulting in robust,
reproducible, mass-produced arrays that can subsequently ``tile"
BICEP-II's focal plane. The tiling with planar arrays is extremely
efficient--- more than five times as many TES detectors can be
placed in the same focal plane area as BICEP's current NTD-Ge
semiconductor array.

BICEP-II will use TES bolometers with integrated planar antennae
\cite{kuo2006SPIE} developed at NASA/JPL and read-out by a
time-domain SQUID multiplexer developed at NIST \cite{dekorte03}.
The ``ultimate" large angular scale B-mode experiment will be a
small array of BICEP-II receivers at the South Pole. This array
would use current technology to probe the GWB if inflation
occurred at, or slightly below, the GUT-scale, thereby testing all
potentially observable models of inflation. All of this can be
accomplished \emph{now} at the South Pole, for approximately $1\%$
of the cost of a similarly capable satellite mission, allowing us
to obtain perhaps the most enigmatic image ever captured: the
birth pangs of the Big Bang!

\section*{Acknowledgments}
This chapter is dedicated to my father, James Ax, whose memory
remains as a continual inspiration. I am grateful for the help and
support of Alexander Polnarev whose insight informed numerous
critical aspects of this work. Evan Bierman, Nathan Miller, and
Thomas Renbarger provided several figures used in this manuscript,
as well as helpful feedback. Insight from Kim Griest, Hans Paar,
and Meir Shimon is gratefully acknowledged. The many successes of
BICEP are attributable to my BICEP team-mates Denis Barkats, Evan
Bierman, Jamie Bock, Cynthia Chiang, Darren Dowell, Lionel Duband,
William Holzapfel, William Jones, John Kovac, Chao-Lin Kuo, Erik
Leitch, Hien Nguyen, Yuki Takahashi, Ki Won Yoon, and Andrew
Lange, BICEP's Principal Investigator. I am grateful for the
assistance of Rob Schluth, Pamela Contractor, and George Ellis for
their hard work as editors on the ``Visions of Discovery" Project.
This work was supported by NSF PECASE Award AST-0548262, the
University of California at San Diego, Caltech President's fund
award PF-471, and the NSF Office of Polar Programs Award
OPP-0230438.





\end{document}